\documentclass[12pt]{iopart}
\usepackage[hyphens]{url}

\usepackage{graphicx}
\usepackage{setstack}
\usepackage{hyperref}
\usepackage{breakurl}
\usepackage[numbers,sort&compress]{natbib}
\usepackage{hyperref}

\begin{document}

\newcommand{\eqref}[1]{(\ref{#1})}
\title[On the theory of the nonlinear Landau damping]{On the theory of the nonlinear Landau damping}

\author{Leon Kos$^{1}$, Ivona Vasileska$^{1}$ and Davy D.~Tskhakaya$^{2}$}

\address{$^1$
 LECAD Laboratory, Faculty of Mechanical Engineering,  University of Ljubljana, SI-1000 Ljubljana, Slovenia\\
 $^2$ Andronikashvili Institute of Physics (TSU), 0177 Tbilisi, Republic of Georgia
}
\ead{leon.kos@lecad.fs.uni-lj.si}
\vspace{10pt}

\begin{abstract}
An exact solution of the collisionless time-dependent Vlasov equation is found 
for the first time. By means of this solution the behavior of the 
Langmuir waves in the nonlinear stage is considered. The analysis is 
restricted by the consideration of the first nonlinear approximation 
keeping the second power of the electric strength. It is shown that in 
general the waves with finite amplitudes are not subject to damping. 
Only in the linear approximation, when the wave amplitude is very small, are 
the waves experiencing damping. It is shown that with the definite 
resonance conditions imposed, the waves become unstable.
\end{abstract}
%
%
\submitto{\jpa}
%
%
%
\section{\label{sec:intro}Introduction}

A large number of papers and textbooks are devoted to the nonlinear theory of 
the Langmuir waves and Landau damping~\cite{landau46_vib_elec_plasma}. 
In addition to the standard way of explanation of physics, a large number of diverse approaches and interpretations have been published~\cite{Weitzner_1963, Weiland_1981, Li_1991, Sagan_1994, Soshnikov_1996, Brodin_1997, Neil_1965, Shatashvili_1982}, which are more advanced than 
the standard one. 

In these papers, the solution of the linearized Vlasov equation is 
used. After finding the zero-order solution of the main equation, the authors 
construct approximations of any higher order~\cite{Tsintsadze_2009, Mazitov1965, bernstein57:_exact_nonlin_plasm_oscil, Drummond_2004, Drummond_2005}. In the present 
paper, however a new approach is presented, allowing to analyze the problem 
self-consistently in an arbitrary order of nonlinear approximation.

The exact solution for Vlasov equation is found for the first time. Using 
this solution, it is shown that the waves with finite amplitude are not 
exposed to damping. Only waves with small amplitudes, when the 
oscillation frequency of captured (in the wave-well) electrons is smaller 
than the damping rate, can damp~\cite{Kadomtzev_1978, Shafranov_1965}. It is found that on the 
fulfillment of the definite resonance condition, the waves with finite 
amplitudes are unstable. For this necessitates, the fulfillment of the
definite resonance conditions, which are similar to conditions with the 
parametric resonance.

\section{\label{sec:2}Exact solution of the Vlasov equation}

We start as usual from the collisionless Vlasov equation and Poisson's 
equation,

\begin{eqnarray}
\label{eq1}
\frac{df}{dt}&=\frac{\partial f}{\partial t}+v\frac{\partial f}{\partial 
z}-E\left( {z,t} \right)\frac{\partial f}{\partial v}=0,\\
\label{eq2}
\frac{\partial E}{\partial z}&=1-n,
\quad
E=-\frac{\partial \phi }{\partial z},
\end{eqnarray}

where the dimensionless values for the time, the coordinate, the velocity 
and the electric potential are used

\begin{equation}
\label{eq3}
\left( {\omega_{pe} t} \right)\to t, \left( {z/\lambda_{De} } \right)\to z,
\left( {v/v_{Te} } \right)\to v, \left( {e\Phi /T_{e} } \right)\to \phi,
\end{equation}

$\omega_{pe} $ is the electron plasma frequency, $\lambda_{De} $ is the 
electron Debye length and $v_{Te} $ is the electron thermal velocity. The 
dimensionless electric field $E$ and the electron number density $n$ are 
defined as follows
\begin{equation}
\label{eq4}
\frac{E}{\sqrt {4\pi \cdot n_{0} T_{e} } }\to E,
\quad
\frac{n}{n_{0} }\to n=\int {dv\cdot f\left( {z,t,v} \right)} .
\end{equation}
Here $T_{e} $ is the electron temperature in the energetic units. It is 
assumed that ions stay in the equilibrium with the density $n_{0} $, which 
results in the first term ("1") on the right-hand side (rhs) of 
Poisson's Eq.~\eqref{eq2}.

The characteristic equations for Eq.~\eqref{eq1} reads
\begin{equation}
\label{eq5}
\frac{dz}{dt}=v,
\quad
\frac{dv}{dt}=-E\left( {z,t} \right).
\end{equation}

For constants of integrals $R$ and $U$ ($dR/dt=0$, $dU/dt=0$ ) we find

\begin{eqnarray}
\label{eq6}
R&=z-\int\limits^{t} {dt'\cdot {H}' \left[ {v+\int\limits_{{t}'}^t 
{d{t}''\cdot E\left\{ {z\left( {{t}''} \right),{t}''} \right\}} } \right]},
\\
\label{eq7}
U&=v+\int\limits^t {d{t}'\cdot E\left\{ {z\left( {{t}'} \right),{t}'} 
\right\}},
\end{eqnarray}

where the functions $z\left( {{t}'} \right)$ are defined 
with the expressions:

\begin{eqnarray}
\label{eq8}
z\left( {{t}'} \right)&=z-\int\limits_{{t}'}^t {d{t}''\cdot {H}'\left[ 
{v+\int\limits_{{t}''}^t {d{t}'''\cdot E\left\{ {z\left( {{t}'''} 
\right),{t}'''} \right\}} } \right]},
\\
\label{eq9}
z\left( {{t}''} \right)&=z-\int\limits_{{t}''}^t {d{t}'''\cdot {H}'\left[ 
{v+\int\limits_{{t}'''}^t {dt^{IV}\cdot E\left\{ {z\left( {t^{IV}} 
\right),t^{IV}} \right\}} } \right]} ,
\\
\label{eq10}
z\left( {{t}'''} \right)&=z-\int\limits_{{t}'''}^t {dt^{IV}\cdot {H}'\left[ 
{v+\int\limits_{t^{IV}}^t {dt^{V}\cdot E\left\{ {z\left( {t^{V}} 
\right),t^{V}} \right\}} } \right]}, \\
&\mathellipsis . \nonumber
\end{eqnarray}

The chain~\eqref{eq8}, \eqref{eq9}, \eqref{eq10}, \textellipsis, can be continued. Here the 
values $z\left( {{t}'} \right)$, $z\left( {{t}''} \right)$, $z\left( {{t}'''} 
\right)$, \textellipsis, must be substituted into the arguments of the 
electric fields' expressions $E\left\{ {z\left( {{t}'} \right),{t}'} 
\right\}$, $E\left\{ {z\left( {{t}''} \right),{t}''} \right\}$, $E\left\{ 
{z\left( {{t}'''} \right),{t}'''} \right\}$, \textellipsis, etc. 
In Eqs.~\eqref{eq6}-\eqref{eq8} the function $H\left( x \right)$ is defined as follows 

\begin{equation}
\label{eq11}
H\left( x \right)=\frac{1}{2}x^{2},
\quad
{H}'\left( x \right)=x.
\end{equation}

In the following the upper dashes in ${H}'\left( x \right)$ will denote the 
derivative with the whole argument of the function $H\left( x \right)$ and 
in ${E}'\left\{ {z\left( t \right),t} \right\}$ will denote a derivative only 
with respect to $z\left( t \right), \quad {E}'\left\{ {z\left( t \right),t} 
\right\}=\partial E\left\{ {z\left( t \right),t} \right\}/\partial 
z\left( t \right)$. The solution of Eq.~\eqref{eq1} can be represented in the form 

\begin{equation}
\label{eq12}
f=f\left\{ {R\left( {v,z,t} \right),U\left( {v,z,t} \right)} \right\}.
\end{equation}
Substituting Eq.~\eqref{eq12} into Eq.~\eqref{eq1}, using the definitions~\eqref{eq6} and~\eqref{eq7}, and 
successively carrying out the derivatives we obtain

\begin{equation}
\label{eq13}
\eqalign{
\frac{df}{dt}=\frac{\partial f}{\partial U}\cdot 
\frac{dU}{dt}+\frac{\partial f}{\partial R}\cdot \frac{dR}{dt}= \cr 
=\left\{ {\frac{\partial f}{\partial U}\cdot \int\limits^t {d{t}'\cdot
{E}'\left\{ {z\left( {{t}'} \right),{t}'} \right\}+\frac{\partial 
f}{\partial R}} } \right\}\cdot \left( {-{H}'\left[ v \right]+v} \right)=0.}
\end{equation}

In Eq.~\eqref{eq13} all other terms cancel each other. According to the second 
relation from~\eqref{eq11} ${H}'\left( v \right)=v$ and hence the solution ~\eqref{eq12} satisfies the kinetic equation~\eqref{eq1}.
Defining the initial distribution function we assume that at the initial 
moment $t_{0} \to -\infty $ the electron distribution function depends only 
on the velocity

\begin{equation}
\label{eq14}
f\left\{ {R\left( {v,z,t_{0} } \right),U\left( {v,z,t_{0} } \right)} 
\right\}=\left. {f_{0} \left( v \right)} \right|_{t_{0\to -\infty } } .
\end{equation}

As $f_{0} \left( v \right)$ we can choose the Maxwell distribution function 
with a normalizing coefficient, $\left( {1/\sqrt {2\pi } } \right)\cdot \exp 
\left( {-v^{2}/2} \right)$. From the relations at the initial time $t_{0} $,

\begin{equation}
\label{eq15}
R\left( {v,z,t_{0} } \right)=R \quad \textrm{and} \quad U\left( {v,z,t_{0} } \right)=U
\end{equation}

and definitions given by Eqs.~\eqref{eq6} and \eqref{eq7} we can find expressions for the velocity $v$ and the coordinate $z$:

\begin{eqnarray}
\label{eq16}
v&=U-\int\limits^{t_{0} } {d{t}'\cdot E\left\{ {z\left( {{t}'} \right),{t}'} 
\right\}},
\\
\label{eq17}
z&=R+\int\limits^{t_{0} } {d{t}'\cdot {H}'} \left[ 
{v+\int\limits_{{t}'}^{t_{0} } {d{t}''\cdot E\left\{ {z\left( {{t}''} 
\right),{t}''} \right\}} } \right].
\end{eqnarray}

Into Eqs.~\eqref{eq16} and \eqref{eq17} the explicit expressions \eqref{eq6} and \eqref{eq7} can be again substituted. Hence using the initial condition \eqref{eq14} for the electron distribution function we find 

\begin{equation}
\label{eq18}
f_{0} =f_{0} \left\{ {v+\int\limits_{t_{0} \to \infty }^t {d{t}'\cdot 
E\left\{ {z\left( {{t}'} \right),{t}'} \right\}} } \right\},
\end{equation}

where values $z\left( {{t}'} \right)$, $z\left( {{t}'''} \right)$, 
\textellipsis, are defined by Eqs.~\eqref{eq8},~\eqref{eq9}, \textellipsis. We can represent Poisson's Eq.~\eqref{eq2}  in the following form,

\begin{equation}
\label{eq19}
\frac{\partial E\left( {z,t} \right)}{\partial z}=1-\int\limits_{-\infty 
}^\infty {dv\cdot f_{0} \left\{ {v+\int\limits_{t_{0} \to -\infty }^t 
{d{t}'\cdot E\left\{ {z\left( {{t}'} \right),{t}'} \right\}} } \right\}} .
\end{equation}

It is convenient to introduce a new variable - s, defined by the relation:

\begin{equation}
\label{eq20}
v+\int\limits_{-\infty }^t {d{t}'\cdot E\left\{ {z\left( {{t}'} 
\right),{t}'} \right\}} =s,
\end{equation}

which allows to simplify the argument of $f_{0} $. Then for the explicit 
expression for the velocity $v$ we have

\begin{equation}
\label{eq21}
\eqalign {v=s-\int\limits_{t_{0} \to -\infty }^t {d{t}'\cdot E\left[ {z\left( {{t}',s} 
\right) } \right] } = s-\int\limits_{t_{0\to .\infty } }^t {d{t}'\cdot E \left[ {z-s\left( {t-{t}'} \right)}+ \right.}\nonumber \cr
+\int\limits_{{t}'}^t {d{t}''\int\limits_{t_{0} }^{{t}''} {d{t}'''\cdot E\left\{ {z-s\left( {t-{t}''} \right)} \right.} } + \nonumber \cr
+\int\limits_{{t}'''}^t {dt^{IV}\int\limits_{t_{0} }^{t^{IV}} {dt^{V}\cdot 
E\left\langle {z-s\left( {t-t^{V}} \right)+....} \right\rangle ,} } \left. 
{{t}'''} \right\} \left. {{t}''} \right],}
\end{equation}

where by analogy to Eqs.~\eqref{eq8}, \eqref{eq9}, \textellipsis, we have introduced the relations: 

\begin{equation}
\label{eq22}
z\left( {{t}',s} \right)=z-s\left\{ {t-{t}'} \right\}+\int\limits_{{t}'}^t 
{d{t}''\int\limits_{t_{0} }^{{t}''} {d{t}'''\cdot E\left[ {z\left( 
{{t}''',s} \right),{t}'''} \right]} },
\end{equation}

\begin{equation}
\label{eq23}
\eqalign{
z\left( {{t}''',s} \right)=z-s\left\{ {t-{t}'''} 
\right\}+\int\limits_{{t}'''}^t {dt^{IV}\int\limits_{t_{0} }^{t^{IV}} 
{dt^{V}\cdot E\left[ {z\left( {t^{V},s} \right),t^{V}} \right]} }\cr
\mathellipsis}
\end{equation}

Poisson's equation can be represented in the form

\begin{eqnarray}
\label{eq24}
\frac{\partial E\left( {z,t} \right)}{\partial z}=1-\int\limits_{-\infty 
}^\infty {ds\cdot \frac{dv\left( s \right)}{ds}f_{0} \left( s \right)}
=1+\int\limits_{-\infty }^\infty {ds\cdot \frac{df_{0} \left( s 
\right)}{ds}\cdot v\left( s \right)}.
\end{eqnarray}

Substituting Eq.~\eqref{eq21} into Eq.~\eqref{eq24} for the Maxwell 
distribution function $f_{0} \left( s \right)$ the first term in~\eqref{eq24} is 
compensated by the positive charge of ions:

\begin{equation}
\label{eq25}
\frac{\partial E}{\partial z}=-\int\limits_{-\infty }^{+\infty } {ds\cdot 
\frac{\partial f_{0} \left( s \right)}{\partial s}\cdot \int\limits_{t_{0} 
\to -\infty }^t {d{t}'\cdot E\left[ {z\left( {{t}'s} \right),{t}'} \right]} 
} .
\end{equation}

Taking the first derivative with time and restricting ourselves to keeping 
the terms up to second powers of the electric strength (such a restriction 
will keep everywhere throughout the calculations) we find

\begin{equation}
\label{eq26}
\eqalign{
\frac{\partial }{\partial t}\frac{\partial E}{\partial 
z}=-\int\limits_{-\infty }^{+\infty } {ds} \cdot \frac{\partial f_{0} \left( 
s \right)}{\partial s}\left\{ {-s\int\limits_{t_{0} \to -\infty }^t 
{d{t}'\cdot {E}'\left[ {z\left( {{t}',s} \right),{t}'} \right]} } \right. \cr
+ \int\limits_{t_{0} \to -\infty }^t {d{t}'\cdot {E}'\left[ {z\left( {{t}',s} 
\right),{t}'} \right]} 
\int\limits_{t_{0} \to -\infty }^t {d{t}'''\cdot E\left[ {z\left( {{t}''',s} 
\right),{t}'''} \right]}
\cr
{-s\int\limits_{t_{0} \to -\infty }^t {d{t}'\cdot {E}'\left[ {z\left( 
{{t}',s} \right),{t}'} \right]}} \cdot \cr
\left.\cdot\int\limits_{{t}'}^t 
{d{t}''\int\limits_{t_{0} \to -\infty }^{{t}''} {d{t}'''\cdot {E}'\left[ 
{z\left( {{t}''',s} \right),{t}'''} \right]} }   \right\}.}
\end{equation}

In the last term in the rhs of this equation we can change the ordering of 
integrals in the following manner

\begin{equation}
\label{eq27}
\int\limits_{t_{0} \to -\infty }^t {d{t}'\int\limits_{{t}'}^t {d{t}''\cdot 
Q\left( {{t}',{t}''} \right)=\int\limits_{t_{0} \to -\infty }^t 
{d{t}''\int\limits_{t_{0} \to -\infty }^{{t}''} {d{t}'} } } } \cdot Q\left( 
{{t}',{t}''} \right).
\end{equation}

Then the second derivative with time of Eq.~\eqref{eq26} gives

\begin{equation}
\label{eq28}
\eqalign{
\frac{\partial^{2}}{\partial t^{2}}\frac{\partial E}{\partial 
z}=-\frac{\partial E}{\partial z}-\int\limits_{-\infty }^{+\infty } {ds} 
\cdot \frac{\partial f_{0} \left( s \right)}{\partial s}
\cdot \left\{ {s^{2}\cdot 
\int\limits_{t_{0} \to -\infty }^t {d{t}'\cdot {E}''\left[ {z\left( {{t}',s} 
\right),{t}'} \right]} } -\right.\cr
-s\cdot \frac{\partial }{\partial z}\cdot \int\limits_{t_{0} \to -\infty }^t 
{d{t}'\cdot {E}'\left\{ {z-s\left( {t-{t}'} \right),{t}'} \right\}} \cr
\cdot \int\limits_{t_{0} \to -\infty }^t {d{t}'''\cdot E\left\{ {z-s\left( 
{t-{t}'''} \right),{t}'''} \right\}}\cr
+\frac{\partial }{\partial t}\cdot \int\limits_{t_{0} \to -\infty }^t 
{d{t}'\cdot {E}'\left\{ {z-s\left( {t-{t}'} \right),{t}'} \right\}}
\cdot \left.\int\limits_{t_{0} \to -\infty }^t {d{t}'''\cdot E\left\{ {z-s\left( 
{t-{t}'''} \right),{t}'''} \right\}}\right\}.}
\end{equation}

In Eq.~\eqref{eq28} we can transform the last term (with the first derivative with 
time) as follows

\begin{equation}
\label{eq29}
\eqalign{
\frac{\partial }{\partial t}\cdot \int\limits_{t_{0} \to -\infty }^t 
{d{t}'\cdot {E}'\left[ {z\left( {{t}',s} \right),{t}'} \right]} 
\int\limits_{t_{0} \to -\infty }^t {d{t}'''\cdot E\left[ {z\left( {{t}''',s} 
\right),{t}'''} \right]} \cong 
\cr
\cong \frac{\partial }{\partial z}\cdot E\left( {z,t} 
\right)\int\limits_{t_{0} \to -\infty }^t {d{t}'\cdot E\left\{ {z-s\left( 
{t-{t}'} \right),{t}'} \right\}}-s\cdot \frac{\partial }{\partial z}\cdot
\cr
\cdot\int\limits_{t_{0} \to -\infty }^t 
{d{t}'\cdot E\left\{ {z-s\left( {t-{t}''} \right),{t}'} \right\}}\cdot \cr
\cdot\int\limits_{t_{0} \to -\infty }^t {d{t}'''\cdot E} \left\{ {z-s\left( 
{t-{t}'''} \right),{t}'''} \right\} ,}
\end{equation}

In the first term on the right-hand side of Eq.~\eqref{eq29}, containing the squared 
electric field, we can, following Eq.~\eqref{eq25}, use a simplified linearized 
expression

\begin{equation}
\label{eq30}
\frac{\partial E}{\partial z}\cong -\int\limits_{-\infty }^{+\infty } 
{ds\cdot \frac{\partial f_{0} \left( s \right)}{\partial 
s}\int\limits_{t_{0} \to -\infty }^t {d{t}'\cdot E\left\{ {z-s\left( 
{t-{t}'} \right),{t}'} \right\}} } .
\end{equation}

Using the relations \eqref{eq29} and \eqref{eq30} the fourth derivative with time of Eq.~\eqref{eq28} we can represent in the form

\begin{equation}
\label{eq31}
\eqalign{
\frac{\partial^{4}E}{\partial t^{4}}+\frac{\partial^{2}E}{\partial 
t^{2}}+3\frac{\partial^{2}E}{\partial z^{2}}+\frac{\partial^{3}}{\partial 
z^{3}}\cdot \cr
\cdot\int\limits_{-\infty }^{+\infty } {ds} \cdot s^{4}\cdot 
\frac{\partial f_{0} \left( s \right)}{\partial s}\int\limits_{t_{0} \to 
-\infty }^t {d{t}'\cdot E\left\{ {z-s\left( {t-{t}'} \right),{t}'} \right\}+} 
\cr
+\frac{\partial }{\partial z}\frac{\partial^{2}}{\partial t^{2}}\left\{ 
{\int\limits_{t_{0} \to -\infty }^t {d{t}'\cdot E\left( {z,{t}'} 
\right)}}\cdot \right. \cr
\left.\cdot \int\limits_{t_{0} \to -\infty }^t {d{t}'\cdot E\left( {z,{t}'} 
\right)}  -\frac{1}{2}\cdot E\left( {z,t} \right)\cdot E\left( {z,t} 
\right) \right\}=0.}
\end{equation}

On obtaining Eq.~\eqref{eq31}, the first derivative $\left( {\partial /\partial z} 
\right)$ in front of every term has been canceled and the explicit 
expressions for the integrals from the Maxwell distribution function 
$-f_{0} \left( s \right)$ is used. 

\section{\label{sec:level3}Waves in the weak nonlinear case}

The first three terms of Eq.~\eqref{eq31} should describe Landau damping 
in the linear approximation. By means of the Fourier expansion of these three terms for the 
electric amplitudes we find 

\begin{equation}
\label{eq32}
E\left( {\omega ,k} \right)=\frac{1}{2\pi }\int\limits_{-\infty }^{+\infty } 
{dt\int\limits_{-\infty }^{+\infty } {dz} \cdot E\left( {z,t} \right)\cdot 
\exp \left\{ {i\left( {\omega t-kz} \right)} \right\}} ,
\end{equation}

which results in the following dispersion relation

\begin{equation}
\label{eq33}
\omega^{4}-\omega^{2}-3k^{2}+\int\limits_{-\infty }^{+\infty } {ds\cdot 
\frac{k^{3}s^{3}}{\omega -ks}\cdot s\frac{\partial f_{0} \left( s 
\right)}{\partial s}} =0.
\end{equation}

A, simple transformation of the last equation gives 

\begin{equation}
\label{eq34}
\omega^{4}+\omega^{3}\int\limits_{-\infty }^{+\infty } {ds\frac{s}{\omega 
-ks}\frac{\partial f_{0} \left( s \right)}{\partial s}=0} .
\end{equation}

After expanding the denominator of \eqref{eq34} in powers of $\left( {ks/\omega } 
\right)$ for the real and the imaginary parts of the frequency, $\omega 
=\omega_{0} +i\gamma $ , we obtain the equalities

\begin{equation}
\label{eq35}
\omega_{0}^{2} =1+3k^{2} \textrm{and} \gamma =-\sqrt {\frac{\pi }{8}} 
\frac{1}{k^{3}}\exp \left\{ {-\frac{1}{2k^{2}}-\frac{3}{2}} \right\}, 
\end{equation}

which determine the frequency and the Landau damping of the high-frequency 
Langmuir waves~\cite{alexandrov84:_princ_plasm_elect}.

In fact Landau damping describes the initial stage of the electrons 
capturing by the wave cavity (the minimum region of the wave), herewith 
the amplitude must be very small, smaller than the value proportional to 
$\gamma $, namely$\sqrt {eE_{0} /T_{e} \cdot k} \ll\gamma /k$~\cite{Kadomtzev_1978} (here 
$E_{0} $ is the amplitude of the wave). In other words, this inequality means 
that the oscillation frequency of captured electrons in the wave-well, must 
be much smaller than the damping rate of the wave $-\gamma $; the electrons 
are pushed by the back-side wall of the wave-cavity and during the 
time-interval of passing the cavity width, the wave should be 
damped~\cite{Kadomtzev_1978, Shafranov_1965}.

Below we consider the case when the inverse inequality is fulfilled, that 
means the predominance of the electrons' oscillation frequency in the well 
over the damping rate $-\gamma$. Then the captured electrons are reflected 
many times from the cavity walls (getting and losing the energy) and on 
average the wave keeps its energy -- hence at $\sqrt{eE_{0} /T_{e} \cdot k} \cdot k\gg\gamma$ the damping of the wave does not take place~\cite{Kadomtzev_1978, Shafranov_1965}. 

To simplify further calculations, it is convenient to introduce the 
following auxiliary value,

\begin{equation}
\label{eq36}
I\left( {z,t} \right)=\int\limits_{t_{0} \to -\infty }^t {d{t}'\cdot E\left( 
{z,{t}'} \right)},
\end{equation}

for which from Eq.~\eqref{eq31} we can obtain the following equation, neglecting the 
term corresponding to the wave damping:

\begin{equation}
\label{eq37}
\frac{\partial^{4}I}{\partial t^{4}}+\frac{\partial^{2}I}{\partial 
t^{2}}+3\frac{\partial^{2}I}{\partial z^{2}}+
\frac{\partial }{\partial z}\frac{\partial }{\partial t}\left\{ {I^{2}\left( 
{z,t} \right)-\frac{1}{2}\left( {\frac{\partial I\left( {z,t} 
\right)}{\partial t}} \right)^{2}} \right\}=0.
\end{equation}

By means of Eq.~\eqref{eq37} we 
can construct the expression for the value $\frac{1}{2}\left( 
{\frac{\partial I\left( {z,t} \right)}{\partial t}} \right)^{2}$. The 
straightforward calculations give

\begin{equation}
\label{eq38}
\eqalign{
\frac{1}{2}\left( {\frac{\partial I\left( {z,t} \right)}{\partial t}} 
\right)^{2}
\cong -\frac{1}{2}I^{2}\left( {z,t} \right)-3\int\limits_{-\infty 
}^t {d{t}'\frac{\partial I\left( {z,{t}'} \right)}{\partial {t}'}} 
\frac{\partial^{2}}{\partial z^{2}}\cdot\cr
\cdot \int\limits_{-\infty }^{{t}'} 
{d{t}''\int\limits_{-\infty }^{{t}''} {d{t}'''\cdot I\left( {z,{t}'''} 
\right)} } -\int\limits_{-\infty }^t {d{t}'\frac{\partial I\left( {z,{t}'} 
\right)}{\partial {t}'}\frac{\partial }{\partial z}}
\cdot\int\limits_{-\infty 
}^{{t}'} {d{t}''\cdot I^{2}\left( {z,{t}''} \right)+}  \mathellipsis .}
\end{equation}

Further we a) hold on to the approximation usually used in the theory of 
Landau damping -- the assumption of the smooth dependence of the 
electric field on the spatial coordinate. Therefore in Eq.~\eqref{eq37} the terms only up to the second derivative with the spatial $z-$coordinate are kept, and b) use 
the relation $\left( {\partial I/\partial t} \right)=E\left( {z,t} 
\right)=-\partial \phi /\partial z$, which follows from Eq.~\eqref{eq36} and 
restrict with the quadratic term $I^{2}$ in the last term in the 
rhs of Eq.~\eqref{eq38}. Assuming the dependence of all unknown values in the 
argument $\xi =z-V\cdot t$, we find that $I\left( \xi \right)=\left( {1/V} 
\right)\cdot \phi \left( \xi \right)$ , where $V$ is the dimensionless 
velocity of the waves, which is assumed to be large, $V\gg 1$. Under 
restrictions a) and b) the last two terms in the rhs of Eq.~\eqref{eq38} give 
negligibly small contributions. Substituting the remaining first term in the 
rhs into Eq.~\eqref{eq37} after two-times integration we obtain

\begin{equation}
\label{eq39}
V^{4}\frac{\partial^{2}\phi }{\partial \xi^{2}}+\left( {V^{2}+3} 
\right)\cdot \phi -\frac{3}{2}\phi^{2}=\frac{1}{2}C_{1} .
\end{equation}

Here the constant $C_{1} $does not depend on time. In the rhs of Eq.~\eqref{eq39} we 
have neglected the term proportional to $\xi $, leading to the nonphysical 
result at $\xi \to \infty $. Multiplying Eq.~\eqref{eq39} by $\left( {\partial \phi 
/\partial \xi } \right)$ after integration we find:

\begin{equation}
\label{eq40}
V^{4}\left( {\frac{\partial \phi }{\partial \xi }} \right)^{2}=\phi 
^{3}-\left( {V^{2}+3} \right)\cdot \phi^{2}+C_{1} \cdot \phi +C_{2} ,
\end{equation}

where the constants of integrations $C_{1} $ and $C_{2} $ can be expressed 
in terms of the minimum $- \quad \phi_{n} $ and the maximum $- \quad \phi_{m} $ 
values of the potential, $\left. {\left( {\partial \phi /\partial \xi } 
\right)} \right|_{\phi_{m} ,\phi_{n} } =0$. As a result we obtain

\begin{equation}
\label{eq41}
V^{4}\left( {\frac{\partial \phi }{\partial \xi }} \right)^{2}=(\phi_{m} 
-\phi )\cdot \left( {\phi_{s} -\phi } \right)\cdot \left( {\phi -\phi_{n} 
} \right).
\end{equation}

The constant $\phi_{s} $ (together with $\phi_{m} $ and $\phi_{n} )$ 
defines the wave velocity $V$:

\begin{equation}
\label{eq42}
V^{2}=\phi_{m} +\phi_{s} +\phi_{n} -3\quad .
\end{equation}

\begin{figure}[htpb]
\centering
\includegraphics{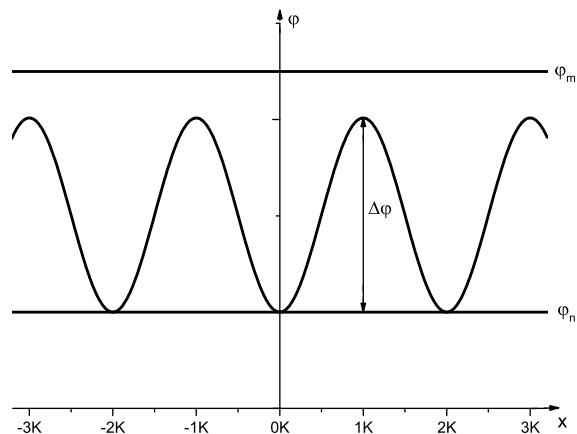}
\caption{Sketch of the dependence of $dn^{2}\left\{ {x,s} \right\}$ on $x$ at $s=const; \Delta \phi =\left( {1-a^{2}} \right)\cdot \left( {\phi_{m} -\phi_{n} } \right)$, $a$ is the minimum value of the curve $dn\left\{ {x,s} \right\}$}
\label{fig:1}
\end{figure}

At $\phi_{m} >\phi_{s} >\phi_{n} $ the solution of Eq.~\eqref{eq41} represents a 
periodic wave, described by the expression (see the Fig.~\ref{fig:1})

\begin{equation}
\label{eq43}
\phi =\phi_{m} -\left( {\phi_{m} -\phi_{n} } \right)\cdot dn^{2}\left\{ 
{x,s} \right\},
\end{equation}

where $dn\left\{ {x,s} \right\}$ is a Jacobi elliptic function with the 
module $-s$~\cite{Jahnke_1945}, 

\begin{equation}
\label{eq44}
dn\left\{ {x,s} \right\}=\frac{\pi }{2\cdot K\left( s \right)}+\frac{2\pi 
}{K\left( s \right)}\sum\nolimits_{n=1}^\infty {\frac{q^{n}}{1+q^{2n}}\cos 
\left\{ {\frac{\pi \cdot n\cdot x}{K\left( s \right)}} \right\}} , 
\end{equation}

\begin{equation}
\label{eq45}
\eqalign{
q=\exp \left[ {-\pi \frac{K\left( {{s}'} \right)}{K\left( s \right)}} 
\right],\quad {s}'=\sqrt {1-s^{2}}, \cr
x=\frac{\sqrt {\phi_{m} -\phi_{n} } }{2\cdot V^{2}}\cdot \xi,
\quad s^{2}=\frac{\phi_{s} -\phi_{n} }{\phi_{m} -\phi_{n}}.}
\end{equation}

The function $dn^{2}\left\{ {x,s} \right\}$ is a periodic function with the 
period $2K\left( s \right)$, where $K\left( s \right)$ is the complete 
elliptic integral of the first kind, 

\begin{equation}
\label{eq46}
K\left( s \right)=\int\limits_0^{\pi /2} {\frac{d\alpha }{\sqrt 
{1-s^{2}\cdot \sin \alpha } }} ,
\end{equation}

therefore the wave length of the periodic solution \eqref{eq43} can be defined 
according to the relation

\begin{equation}
\label{eq47}
\lambda =\frac{4}{\sqrt {\phi_{m} -\phi_{n} } }\cdot V^{2}K\left( s 
\right)
\end{equation}

\section{Instability of waves}
We can now analyze the stability of the waves \eqref{eq43}. Starting from Eq.~\eqref{eq39} 
we introduce the potential perturbation $\delta \phi $ according to the 
relation $\phi \to \phi +\delta \phi $ . For the unperturbed part of the 
potential$-\phi $ remains the expression \eqref{eq43} and for $\delta \phi $ we 
obtain

\begin{equation}
\label{eq48}
\frac{\partial^{2}\delta \phi }{\partial \xi^{2}}+\frac{1}{V^{2}}\left[ 
{1+\frac{3-\phi }{V^{2}}+\frac{3}{V^{2}}\left\{ {1-dn^{2}\left( {z,s^{2}} 
\right)} \right\}} \right]\cdot \delta \phi =0.
\end{equation}

An explicit solution of this equation with the periodic coefficient can be 
found applying a Hill's method~\cite{Whittaker_1973}. This method is rather cumbersome with 
the long and difficult calculations. Here we simplify the equation recalling 
the condition used above, $V \gg 1$, and assuming that 
the parameter $s$ is small, $s \ll 1$. Then Eq.~\eqref{eq48} can be reduced to 
Mathieu's equation~\cite{Whittaker_1973, landau05:_mech}, which describes the phenomenon known as a parametric resonance,

\begin{equation}
\label{eq49}
\frac{\partial^{2}\delta \phi }{\partial \xi^{2}}+\frac{1}{V^{2}}\left\{ 
{1+h\cdot \cos k\xi } \right\}\cdot \delta \phi =0 ,
\end{equation}

where $h\cong \frac{12}{V^{2}}\frac{\phi_{m} -\phi_{n} }{Ch\pi }$ and 
$k\cong \frac{\sqrt {\phi_{m} -\phi_{n} } }{V^{2}}$. Applying a standard 
method at the fulfilment of the resonance condition $k=2/V$ Eq.~\eqref{eq49} gives 
the following expression for the rate of the instability $\bar{{\gamma 
}}=\frac{1}{4}\frac{h}{V}$~\cite{landau05:_mech}.

\section{Summary}

Finding the exact analytic solution for the collisionless Vlasov's equation, 
the nonlinear stage of the Langmuir waves is analyzed in the first 
non-vanishing (quadratic) nonlinear approximation. The Langmuir waves with 
the finite amplitude, and with the oscillation frequency (of electrons in 
the wave-well) larger than the damping rate (found in the linear 
approximation), do not damp and tend to keep the periodic structure. On 
the fulfilment of the definite resonance conditions, the waves are unstable. 
The finding of the corresponding rate is quite similar to the procedure applied in
the investigation of the parametric instability.

\section{Acknowledgment}

This work has been partially supported by the grants P2-0073 and P2-0256 of the Slovenian Research Agency. IV would like to acknowledge the financial support from the EUROfusion Consortium under grant agreement No 633053  (WP-EDU). The views and opinions expressed herein do not necessarily reflect those of the European Commission.




\newcommand{\newblock}{}
\bibliographystyle{unsrtnat}
\providecommand{\noopsort}[1]{}\providecommand{\singleletter}[1]{#1}%

\end{document}